\documentclass[twocolumn, amsmath, amssymb, aps, superscriptaddress, floatfix]{revtex4}
\pdfoutput=1

\usepackage{color}
\usepackage{graphicx}
\usepackage{xspace}

\newcommand{\x}{\,$\times$\,}

\begin{document}
 
\title{X-ray imaging of the dynamic magnetic vortex core deformation}

\author{A. Vansteenkiste}
\affiliation{
  Department of Subatomic and Radiation Physics, Ghent University, Proeftuinstraat 86, 9000 Gent, Belgium.
}

\author{K. W. Chou}
\affiliation{
  Advanced Light Source, LBNL, One Cyclotron Road, 94720 Berkeley, CA, USA.
}

\author{M. Weigand}
\affiliation{
  Max-Planck-Institut f{\"u}r Metallforschung, Heisenbergstr. 3, 70596 Stuttgart, Germany.
}

\author{M. Curcic}
\affiliation{
  Max-Planck-Institut f{\"u}r Metallforschung, Heisenbergstr. 3, 70596 Stuttgart, Germany.
}

\author{V. Sackmann}
\affiliation{
  Max-Planck-Institut f{\"u}r Metallforschung, Heisenbergstr. 3, 70596 Stuttgart, Germany.
}

\author{H. Stoll}
\affiliation{
  Max-Planck-Institut f{\"u}r Metallforschung, Heisenbergstr. 3, 70596 Stuttgart, Germany.
}

\author{T. Tyliszczak}
\affiliation{
  Advanced Light Source, LBNL, One Cyclotron Road, 94720 Berkeley, CA, USA.
}

\author{G. Woltersdorf}
\affiliation{
  Institut f{\"u}r Experimentelle und Angewandte Physik, Universit\"at Regensburg, 93040 Regensburg, Germany.
}

\author{C. H. Back}
\affiliation{
  Institut f{\"u}r Experimentelle und Angewandte Physik, Universit\"at Regensburg, 93040 Regensburg, Germany.
}

\author{G. Sch{\"u}tz}
\affiliation{
  Max-Planck-Institut f{\"u}r Metallforschung, Heisenbergstr. 3, 70596 Stuttgart, Germany.
}

\author{B. {Van Waeyenberge}}
\affiliation{
  Max-Planck-Institut f{\"u}r Metallforschung, Heisenbergstr. 3, 70596 Stuttgart, Germany.
}

\date{\today}


\begin{abstract}
Magnetic platelets with a vortex configuration are attracting considerable attention. The discovery that excitation with small in-plane magnetic fields \cite{vanwaeyenberge06} or spin polarised currents \cite{yamada07} can switch the polarisation of the vortex core did not only open the possibility of using such systems in magnetic memories, but also initiated the fundamental investigation of the core switching mechanism itself. Micromagnetic models predict that the switching is mediated by a vortex-antivortex pair, nucleated in a dynamically induced vortex core deformation. This theoretical framework also predicts a critical core velocity, above which switching occurs \cite{yamada07, guslienko08}. Although this model is extensively studied and generally accepted \cite{kim07, kim08, gliga08a, gliga08b, gliga08c, guslienko08, liu07, liu07b, lee07, xiao07, kravchuk07, hertel07}, experimental support has been lacking until now. In this work, we have used high-resolution time-resolved X-ray microscopy to study the detailed dynamics of vortices. We reveal the dynamic vortex core deformation preceding the core switching.  In addition, the threshold velocity is directly measured, allowing for a quantitative comparison with micromagnetic models.
\end{abstract}

\maketitle

A magnetic vortex appears as the ground state in soft magnetic, micron-sized structures with adequate dimensions. This configuration is characterised by an in-plane curling magnetisation, and an out-of-plane vortex core at the centre of
the structure \cite{feldtkeller65, raabe00}. The diameter of the vortex core is typically only 10 -- 25\,nm \cite{wachowiak02, chou07}, depending on the material and thickness.\\ 

It was shown that a low-field excitation of the so-called vortex gyration mode \cite{huber82, argyle84, guslienko02} can switch the out-of-plane polarisation of the vortex core \cite{vanwaeyenberge06}. This was experimentally observed by determining the vortex core polarisation before and after the application of short bursts of an alternating magnetic field \cite{vanwaeyenberge06}. The dynamic process behind the switching could not be inferred from this experiment. However, micromagnetic modelling showed that near a moving vortex core, a region appears where the magnetisation acquires an out-of-plane component opposing the vortex core polarisation. If this so-called vortex core deformation becomes so strong that it points fully out of the sample plane, a vortex-antivortex pair is nucleated. At this point, the switching is initiated, as the antivortex rapidly annihilates with the original vortex, leaving behind only the newly created vortex with an opposite core polarisation \cite{vanwaeyenberge06, hertel07, gaididei08}. This annihilation process involves a magnetic singularity, which is necessary for the switching \cite{thiaville03}.\\

Apart from micromagnetic simulations, the dynamic core deformation had already been included in theoretical calculations by Novosad \textit{et al.} \cite{novosad05}. Its origin and relevance for the switching process were investigated by Yamada \textit{et al.} \cite{yamada07} and by Guslienko \textit{et al.} \cite{guslienko08}. These  authors show that near a moving vortex,
a strong out-of-plane ``kinetic'' term in the effective field appears. It is this field that pushes the magnetisation towards the opposite direction of the core polarisation, causing the dynamic deformation.\\

Since the effective out-of-plane field is proportional to the velocity of the vortex movement \cite{yamada07}, a critical velocity must exist at which a vortex-antivortex pair is created and core switching will occur \cite{yamada07}. The critical velocity is predicted to be independent of the strength and type of the applied excitation and to only depend on the exchange parameter $A$ \cite{guslienko08}. Calculated critical velocities for typical Permalloy systems range from 250\,m/s \cite{yamada07} to 320\,m/s \cite{guslienko08}.\\

We have used time-resolved magnetic scanning transmission X-ray microscopy (STXM) to image the detailed dynamics of the out-of-plane component of the magnetisation in Permalloy nanostructures. A spatial resolution of about 30\,nm and a temporal resolution of 100\,ps could be reached (see Methods). Micromagnetic simulations (see Methods) yield a vortex core diameter of 25.4\,nm (full width half maximum, averaged over the layer thickness) for the investigated structure. Although this is smaller than the lateral resolution of the microscope, it can still be resolved but appears smeared out \cite{chou07}.\\

\begin{figure}[!htb]
\centering
\includegraphics[width=1\linewidth]{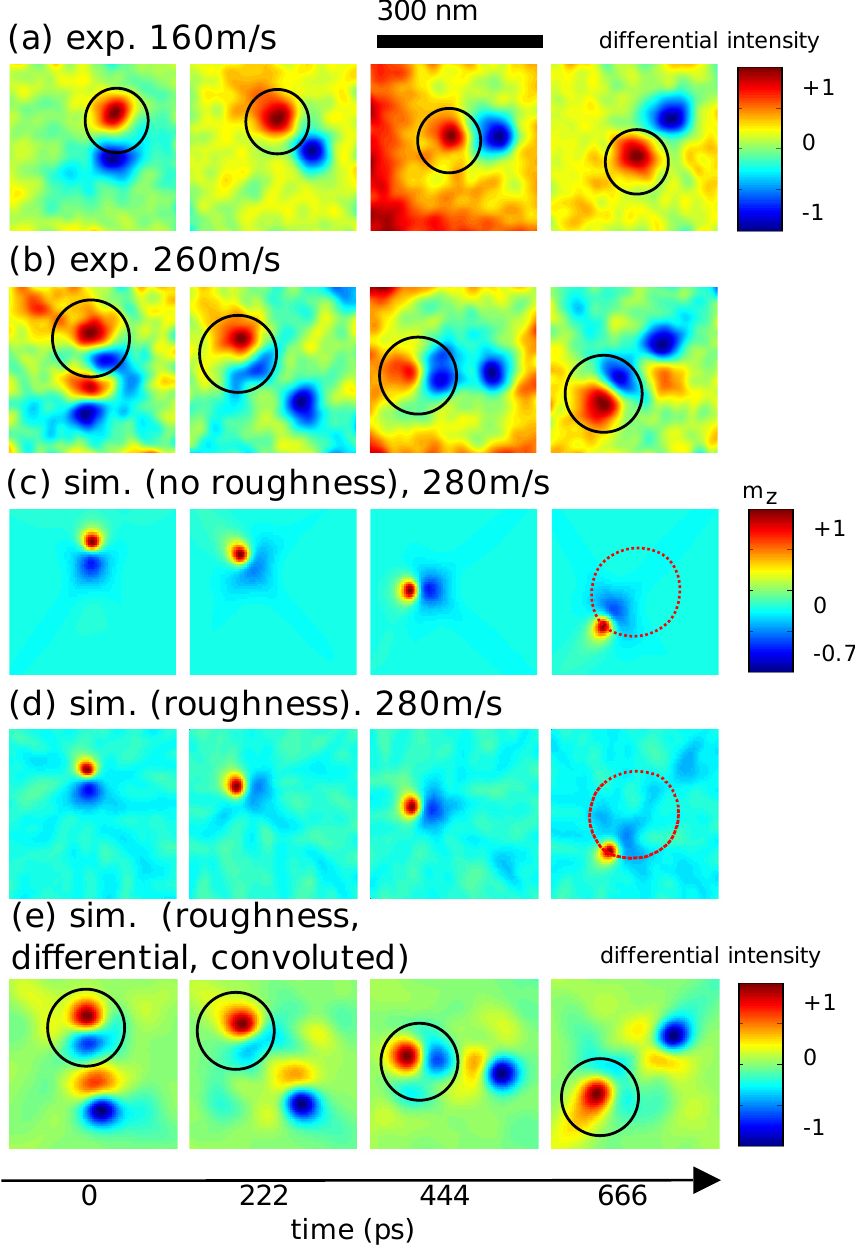}
\caption{\label{fig1} Experimental and simulated images of the time-dependent out-of-plane component of the magnetisation of a vortex structure. (a) Differential STXM images  under a continuous 0.32\,mT rf excitation. Red/blue corresponds to the magnetisation pointing up/down. The vortex core can be seen as a red spot, marked by a circle (the second spot originates from the subtraction of the reference image). (b): Images of the same sample, but with a 0.89\,mT excitation. The dynamic vortex core deformation is now visible as an additional spot near the vortex core. (c-d) Simulation of steady-state vortex gyration in a Permalloy structure under similar conditions as the experiment, with and without sample roughness (see Methods). The core trajectory is shown with a dashed line. (e): Differential images generated from the simulation(d), convoluted with the experimental lateral resolution. These images can be directly compared to the experimental images in (b).}
\end{figure}

 Fig. \ref{fig1}a-b shows STXM images of the centre portion of a 500\,nm\, $\times$\,500\,nm\,$\times$\,50\,nm Permalloy square structure. The structure is excited with a continuous in-plane rf magnetic field oscillating at a frequency of 562.5 MHz, close to the eigenfrequency of its gyrotropic mode \cite{vansteenkiste08}. Under these conditions, a steady-state gyration of the vortex is obtained. In order to remove the non-magnetic contrast contributions, reference images recorded at a 180$^\circ$ phase shift of the rf excitation are subtracted from the respective images. In these so-called differential images, the vortex core appears twice: once from the original image, and once with inverted contrast from the reference image. The positions of these two vortex core images are mirrored
with respect to the structure-centre and appear well-separated when the gyration radius is larger than the lateral resolution of the microscope --- about 30\,nm.\\

The stroboscopic images were recorded at eight phases of the gyration cycle, with time intervals of 222\,ps between them. The last four images were then subtracted from the respective first four, and
the resulting four differential images are shown. In  Fig. \ref{fig1}a, the vortex core points up and appears as a red spot, marked
with a circle. Note that the vortex core gyrates counterclockwise, as dictated by its polarisation \cite{huber82}. The blue spot in the image originates from the subtraction of the reference image, and corresponds to a negative image of the core appearing with a 180$^\circ$ phase shift in the gyration cycle. The amplitude of the excitation is here 0.32\,mT and causes the core to
gyrate with an average velocity of (160\,$\pm$\,20)\,m/s. This velocity was determined from the average radius $r$ of the
gyration, since the angular velocity $2\pi f $ is fixed and $v=2\pi f r$. {The trajectory was thus assumed to be approximately circular, which is verified by the simulations in Fig. \ref{fig1}c-d.}\\

\begin{figure}
 \begin{centering}
  \includegraphics[width=0.9\linewidth]{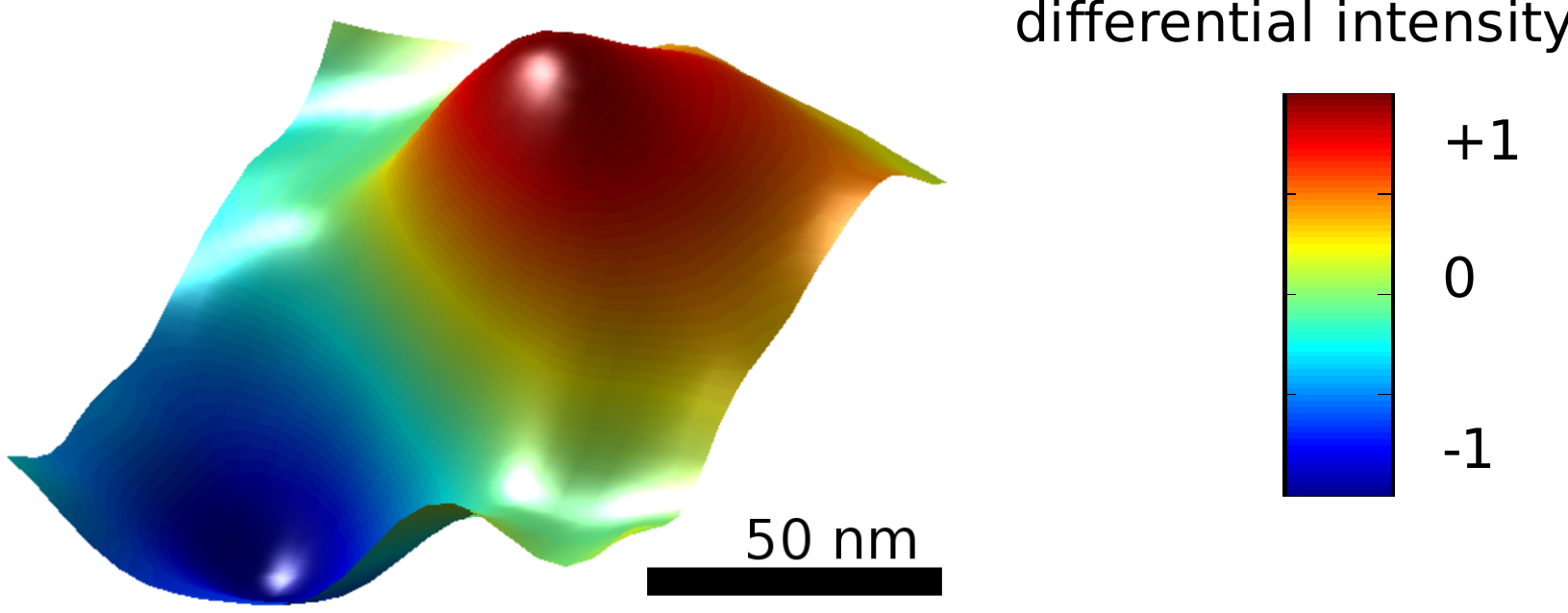}
 \end{centering}
 \caption{Three-dimensional representation of the experimentally observed vortex core profile, generated from the marked area in the last frame of Fig. \ref{fig1}b. The differential intensity is proportional to the out-of-plane magnetisation. Although the features are smaller than the resolution of the microscope and are therefore significantly smeared out, the bipolar nature of the vortex core profile can still be clearly observed. \label{fig2}}\clearpage
\end{figure}

When the excitation amplitude is increased, a larger gyration amplitude, and consequently a larger velocity, can be observed.
At 0.89\,mT (Fig. \ref{fig1}b) the average gyration velocity is (260\,$\pm$\,20)\,m/s. This excitation is already very close to the switching threshold. When the field was increased to 1\,mT, the vortex core polarisation was found to be switching back and forth.  Further investigation of the images recorded just below the switching threshold reveals an additional spot near the vortex core (also shown in Fig. \ref{fig2}). Having a magnetisation opposite to the core, this spot is identified as the dynamic vortex core deformation. This deformation was predicted as the nucleation site of the vortex-antivortex pair when the gyration velocity reaches the threshold for core switching \cite{yamada07, guslienko08}.\\

\begin{figure}[!htb]
\centering
\includegraphics[width=0.9\linewidth]{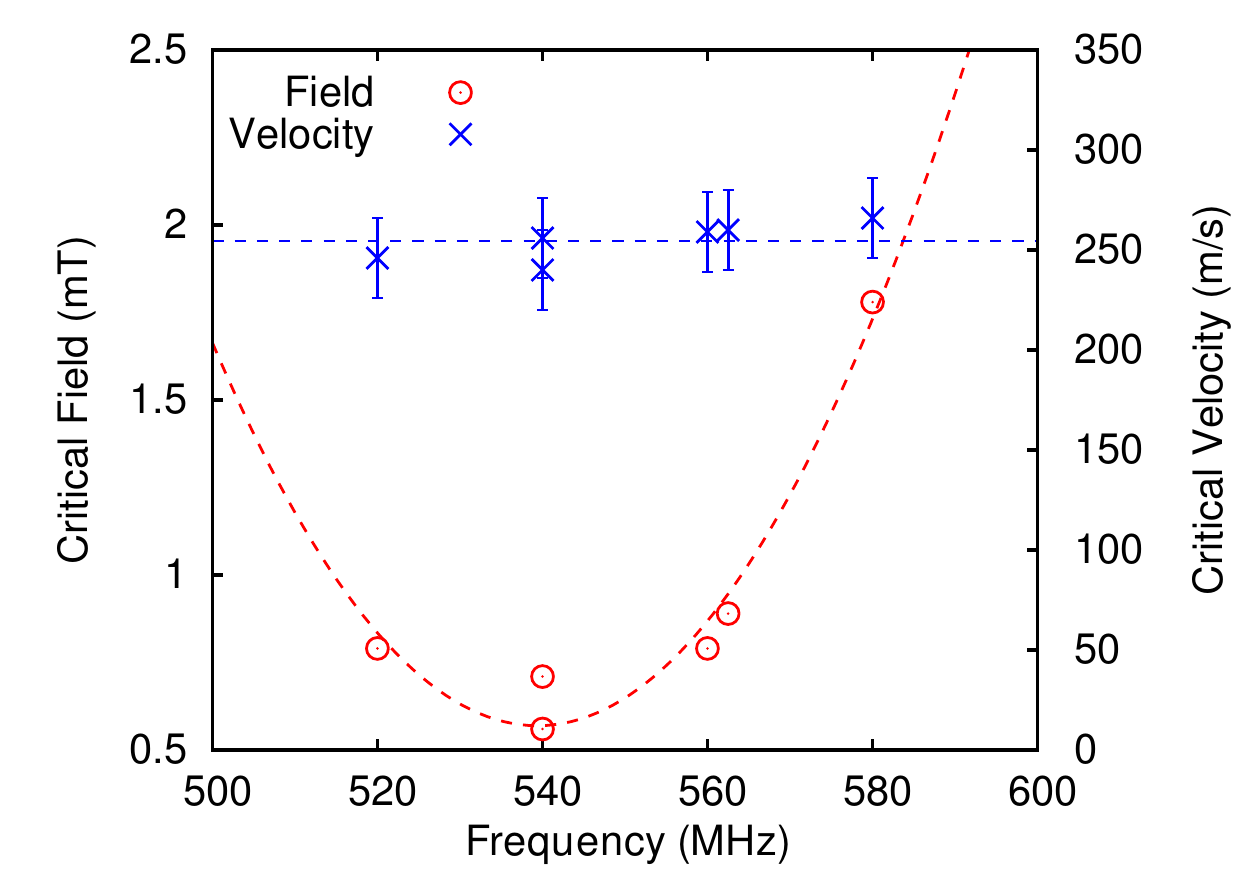}
\caption{\label{fig3} Magnetic field amplitudes $B_{0,\mathrm{thr}}$ and average vortex velocities $v_{\mathrm{thr}}$ just below the threshold for core switching, determined for different excitation frequencies $f$ near the gyrotropic resonance of three 500\,nm\,$\times$\,500\,nm\,$\times$\,50\,nm Permalloy samples. The critical field required for switching (red points) varies by more than a factor of 2 when the excitation frequency is detuned away from the resonance. The critical velocity (blue points), however, remains constant within the experimental error (the standard deviation of the velocities, determined at 8 different phases of the movement, rounded up). The dashed curves are a guide to the eye.}\clearpage
\end{figure}

Besides the direct observation of the dynamic vortex core deformation, these experiments also allow to determine the critical velocity for core switching. This was done by increasing the excitation in small steps of maximum 12\,\% until vortex core reversal was observed. The critical switching velocity was estimated by determining the maximum core velocity, right below the switching threshold \cite{yamada07}. This procedure was repeated when the excitation frequency was detuned, away from the gyrotropic resonance where higher excitation amplitudes are necessary to achieve switching. The highest measured velocities of various samples at different excitation amplitudes are shown in Fig. \ref{fig3}. Even when the magnetic field amplitude required for switching varied by more than a factor of two, the  core velocity just below the threshold was found to be constant within the experimental error. The vortex core velocity thus appears indeed to be the critical switching parameter, regardless of excitation amplitude and frequency.\\\\


The experimental results could be well reproduced by micromagnetic simulations. A Permalloy platelet with the same dimensions as the experimentally investigated ones was simulated. In Fig. \ref{fig1}c, the simulated out-of-plane component of the magnetisation is shown during a steady-state gyration just below the switching threshold. The vortex core deformation can be distinguished as a spot with opposite polarisation next to the vortex core. The strength of the deformation is not entirely constant, but its fluctuations follow the 4-fold symmetry of the square sample. In order to reproduce the experimental results as well as possible, simulations including a small surface roughness (see Methods) were performed as well (Fig. \ref{fig1}d). In our evaporated thin films, such surface roughness is inevitable. The roughness did not change the underlying switching mechanism itself, but was found to only induce some additional fluctuations in the deformation profile as well as some additional out-of-plane magnetisation components, as is also seen in the experimental images.\\

To illustrate the correspondence between the simulations with roughness and the experiment, a series of differential images was generated from these simulations (Fig. \ref{fig1}d). These are obtained via the same procedure as the experimental differential images (from each image, the image corresponding to a 180$^\circ$ phase shift of the rf excitation has been subtracted), and have been convoluted with a gaussian function corresponding to the lateral resolution of the microscope. The position and size of the deformation in these images can be seen to correspond well to the experimental data in Fig. \ref{fig1}b.\\

From the simulations, an instantaneous critical velocity of 300\,m/s was found, which is well in the range of the predicted velocities \cite{yamada07, guslienko08}. The vortex core velocity was however found to fluctuate along the trajectory. This is not only due to the repetitive acceleration by the rf field, but also due to the surface roughness, which causes some additional fluctuations. Therefore, the \emph{average} velocity over one period was slightly lower than the peak velocity: about 280\,m/s. This is in good agreement with the experimental average velocities, considering that the experimental values are lower limits of the actual switching threshold.\\

{In conclusion, we have experimentally observed the dynamic vortex core deformation and threshold velocity, providing the first strong experimental support for the microscopic switching model via vortex-antivortex creation and annihilation. This is the first time that ``internal'' dynamics of the vortex core could be imaged by time-resolved X-ray microscopy. Further improvement of the spatial and temporal resolution may even open the possibility to observe the vortex-antivortex pair creation and annihilation itself.}

\section*{Methods}

\subsection*{Experiments}

High-resolution images of the vortex core dynamics in Permalloy nanostructures were recored using the scanning transmission X-ray microscope at the Advanced Light Source (ALS, beamline 11.0.2) \cite{kilcoyne03}. This microscope has a resolution of about 30\, nm. Since the pixel size was only 5\,nm, the presented images could be slightly smoothed to reduce the noise, without loosing significant information (no such smoothing was applied during the quantitative analysis). The investigated nanostructures were square-shaped and patterned from an evaporated Permalloy film on top of a 2.5\,$\mu$m wide and 150\,nm thick Cu stripline used for the magnetic excitation. The magnetisation is imaged using the X-ray magnetic circular dichroism (XMCD) effect \cite{schutz87}. By orienting the sample plane perpendicular to the photon beam, the out-of-plane component of the magnetisation is measured, allowing direct imaging of the vortex core \cite{chou07}. An rf current is transmitted through the stripline underneath the magnetic structures and induces an in-plane rf magnetic field $B(t)=B_0\sin\left(2\pi f t\right)$. By synchronising the excitation with the X-ray flashes of the synchrotron, stroboscopic images of the moving vortex core can be recorded. The time resolution of these images is limited by the width of the photon flashes to about 100\,ps. \\

\subsection*{Simulations}

A 500\,nm\,$\times$\,500\,nm\,$\times$\,50\,nm platelet was simulated with a cell size of 3.9\,nm\,$\times$\,3.9\,nm\,$\times$\,$\pm$12.5\,nm. Typical material parameters for Permalloy were used: saturation magnetisation $M_s$=800\x10$^3$\,A/m, exchange constant $A$=13\x 10$^{-12}$\,J/m, damping parameter $\alpha$=0.01,
anisotropy constant $K_1$=0. An excitation of 1.9\,mT at 562.5\,MHz was applied, which is just below the switching threshold. The \textsc{amumag} micromagnetic package, an in-house developed 3D finite-element code, was used to solve the Landau-Lifshitz equation \cite{landau35}. The source code is available at {http://code.google.com/p/amumag}. The Fast Multipole Method \cite{visscher04} was employed to calculate the magnetostatic fields, assuming constant magnetisation in each of the hexahedral cells. The simulations did not include thermal fluctuations. In order to match the realistic sample geometry as close as possible, a small random surface roughness was introduced (rms amplitude $\approx$ 5\,nm,  correlation length $\approx$ 20\,nm).

\bibliography{biblio}

\begin{thebibliography}{29}
\expandafter\ifx\csname natexlab\endcsname\relax\def\natexlab#1{#1}\fi
\expandafter\ifx\csname bibnamefont\endcsname\relax
  \def\bibnamefont#1{#1}\fi
\expandafter\ifx\csname bibfnamefont\endcsname\relax
  \def\bibfnamefont#1{#1}\fi
\expandafter\ifx\csname citenamefont\endcsname\relax
  \def\citenamefont#1{#1}\fi
\expandafter\ifx\csname url\endcsname\relax
  \def\url#1{\texttt{#1}}\fi
\expandafter\ifx\csname urlprefix\endcsname\relax\def\urlprefix{URL }\fi
\providecommand{\bibinfo}[2]{#2}
\providecommand{\eprint}[2][]{\url{#2}}

\bibitem[{\citenamefont{{Van Waeyenberge} et~al.}(2006)\citenamefont{{Van
  Waeyenberge}, Puzic, Stoll, Chou, Tyliszczak, Hertel, Fahnle, Bruckl, Rott,
  Reiss et~al.}}]{vanwaeyenberge06}
\bibinfo{author}{\bibfnamefont{B.}~\bibnamefont{{Van Waeyenberge}}},
  \bibinfo{author}{\bibfnamefont{A.}~\bibnamefont{Puzic}},
  \bibinfo{author}{\bibfnamefont{H.}~\bibnamefont{Stoll}},
  \bibinfo{author}{\bibfnamefont{K.~W.} \bibnamefont{Chou}},
  \bibinfo{author}{\bibfnamefont{T.}~\bibnamefont{Tyliszczak}},
  \bibinfo{author}{\bibfnamefont{R.}~\bibnamefont{Hertel}},
  \bibinfo{author}{\bibfnamefont{M.}~\bibnamefont{Fahnle}},
  \bibinfo{author}{\bibfnamefont{H.}~\bibnamefont{Bruckl}},
  \bibinfo{author}{\bibfnamefont{K.}~\bibnamefont{Rott}},
  \bibinfo{author}{\bibfnamefont{G.}~\bibnamefont{Reiss}},
  \bibnamefont{et~al.}, \bibinfo{journal}{Nature}
  \textbf{\bibinfo{volume}{444}}, \bibinfo{pages}{461} (\bibinfo{year}{2006}).

\bibitem[{\citenamefont{Yamada et~al.}(2007)\citenamefont{Yamada, Kasai,
  Nakatani, Kobayashi, Kohno, Thiaville, and Ono}}]{yamada07}
\bibinfo{author}{\bibfnamefont{K.}~\bibnamefont{Yamada}},
  \bibinfo{author}{\bibfnamefont{S.}~\bibnamefont{Kasai}},
  \bibinfo{author}{\bibfnamefont{Y.}~\bibnamefont{Nakatani}},
  \bibinfo{author}{\bibfnamefont{K.}~\bibnamefont{Kobayashi}},
  \bibinfo{author}{\bibfnamefont{H.}~\bibnamefont{Kohno}},
  \bibinfo{author}{\bibfnamefont{A.}~\bibnamefont{Thiaville}},
  \bibnamefont{and} \bibinfo{author}{\bibfnamefont{T.}~\bibnamefont{Ono}},
  \bibinfo{journal}{Nat. Mater.} \textbf{\bibinfo{volume}{6}},
  \bibinfo{pages}{269} (\bibinfo{year}{2007}).

\bibitem[{\citenamefont{Guslienko et~al.}(2008)\citenamefont{Guslienko, Lee,
  and Kim}}]{guslienko08}
\bibinfo{author}{\bibfnamefont{K.~Y.} \bibnamefont{Guslienko}},
  \bibinfo{author}{\bibfnamefont{K.~S.} \bibnamefont{Lee}}, \bibnamefont{and}
  \bibinfo{author}{\bibfnamefont{S.~K.} \bibnamefont{Kim}},
  \bibinfo{journal}{Phys. Rev. Lett.} \textbf{\bibinfo{volume}{100}},
  \bibinfo{pages}{027203} (\bibinfo{year}{2008}).

\bibitem[{\citenamefont{Kim et~al.}(2007)\citenamefont{Kim, Choi, Lee,
  Guslienko, and Jeong}}]{kim07}
\bibinfo{author}{\bibfnamefont{S.~K.} \bibnamefont{Kim}},
  \bibinfo{author}{\bibfnamefont{Y.~S.} \bibnamefont{Choi}},
  \bibinfo{author}{\bibfnamefont{K.~S.} \bibnamefont{Lee}},
  \bibinfo{author}{\bibfnamefont{K.~Y.} \bibnamefont{Guslienko}},
  \bibnamefont{and} \bibinfo{author}{\bibfnamefont{D.~E.} \bibnamefont{Jeong}},
  \bibinfo{journal}{Appl. Phys. Lett.} \textbf{\bibinfo{volume}{91}},
  \bibinfo{pages}{082506} (\bibinfo{year}{2007}).

\bibitem[{\citenamefont{Kim et~al.}(2008)\citenamefont{Kim, Lee, Choi,
  Guslienko, and Lee}}]{kim08}
\bibinfo{author}{\bibfnamefont{S.~K.} \bibnamefont{Kim}},
  \bibinfo{author}{\bibfnamefont{J.~Y.} \bibnamefont{Lee}},
  \bibinfo{author}{\bibfnamefont{Y.~S.} \bibnamefont{Choi}},
  \bibinfo{author}{\bibfnamefont{K.~Y.} \bibnamefont{Guslienko}},
  \bibnamefont{and} \bibinfo{author}{\bibfnamefont{K.~S.} \bibnamefont{Lee}},
  \bibinfo{journal}{Appl. Phys. Lett.} \textbf{\bibinfo{volume}{93}},
  \bibinfo{pages}{052503} (\bibinfo{year}{2008}).

\bibitem[{\citenamefont{Gliga et~al.}(2008{\natexlab{a}})\citenamefont{Gliga,
  Yan, Hertel, and Schneider}}]{gliga08a}
\bibinfo{author}{\bibfnamefont{S.}~\bibnamefont{Gliga}},
  \bibinfo{author}{\bibfnamefont{M.}~\bibnamefont{Yan}},
  \bibinfo{author}{\bibfnamefont{R.}~\bibnamefont{Hertel}}, \bibnamefont{and}
  \bibinfo{author}{\bibfnamefont{C.~M.} \bibnamefont{Schneider}},
  \bibinfo{journal}{Phys. Rev. B} \textbf{\bibinfo{volume}{77}},
  \bibinfo{pages}{060404} (\bibinfo{year}{2008}{\natexlab{a}}).

\bibitem[{\citenamefont{Gliga et~al.}(2008{\natexlab{b}})\citenamefont{Gliga,
  Hertel, and Schneider}}]{gliga08b}
\bibinfo{author}{\bibfnamefont{S.}~\bibnamefont{Gliga}},
  \bibinfo{author}{\bibfnamefont{R.}~\bibnamefont{Hertel}}, \bibnamefont{and}
  \bibinfo{author}{\bibfnamefont{C.~M.} \bibnamefont{Schneider}},
  \bibinfo{journal}{Physica B-condensed Matter} \textbf{\bibinfo{volume}{403}},
  \bibinfo{pages}{334} (\bibinfo{year}{2008}{\natexlab{b}}).

\bibitem[{\citenamefont{Gliga et~al.}(2008{\natexlab{c}})\citenamefont{Gliga,
  Hertel, and Schneider}}]{gliga08c}
\bibinfo{author}{\bibfnamefont{S.}~\bibnamefont{Gliga}},
  \bibinfo{author}{\bibfnamefont{R.}~\bibnamefont{Hertel}}, \bibnamefont{and}
  \bibinfo{author}{\bibfnamefont{C.~M.} \bibnamefont{Schneider}},
  \bibinfo{journal}{J. Appl. Phys.} \textbf{\bibinfo{volume}{103}},
  \bibinfo{pages}{07B115} (\bibinfo{year}{2008}{\natexlab{c}}).

\bibitem[{\citenamefont{Liu et~al.}(2007{\natexlab{a}})\citenamefont{Liu, He,
  and Zhang}}]{liu07}
\bibinfo{author}{\bibfnamefont{Y.~W.} \bibnamefont{Liu}},
  \bibinfo{author}{\bibfnamefont{H.}~\bibnamefont{He}}, \bibnamefont{and}
  \bibinfo{author}{\bibfnamefont{Z.~Z.} \bibnamefont{Zhang}},
  \bibinfo{journal}{Appl. Phys. Lett.} \textbf{\bibinfo{volume}{91}},
  \bibinfo{pages}{242501} (\bibinfo{year}{2007}{\natexlab{a}}).

\bibitem[{\citenamefont{Liu et~al.}(2007{\natexlab{b}})\citenamefont{Liu,
  Gliga, Hertel, and Schneider}}]{liu07b}
\bibinfo{author}{\bibfnamefont{Y.}~\bibnamefont{Liu}},
  \bibinfo{author}{\bibfnamefont{S.}~\bibnamefont{Gliga}},
  \bibinfo{author}{\bibfnamefont{R.}~\bibnamefont{Hertel}}, \bibnamefont{and}
  \bibinfo{author}{\bibfnamefont{C.~M.} \bibnamefont{Schneider}},
  \bibinfo{journal}{Appl. Phys. Lett.} \textbf{\bibinfo{volume}{91}},
  \bibinfo{pages}{112501} (\bibinfo{year}{2007}{\natexlab{b}}).

\bibitem[{\citenamefont{Lee et~al.}(2007)\citenamefont{Lee, Guslienko, Lee, and
  Kim}}]{lee07}
\bibinfo{author}{\bibfnamefont{K.~S.} \bibnamefont{Lee}},
  \bibinfo{author}{\bibfnamefont{K.~Y.} \bibnamefont{Guslienko}},
  \bibinfo{author}{\bibfnamefont{J.~Y.} \bibnamefont{Lee}}, \bibnamefont{and}
  \bibinfo{author}{\bibfnamefont{S.~K.} \bibnamefont{Kim}},
  \bibinfo{journal}{Phys. Rev. B} \textbf{\bibinfo{volume}{76}},
  \bibinfo{pages}{174410} (\bibinfo{year}{2007}).

\bibitem[{\citenamefont{Xiao et~al.}(2007)\citenamefont{Xiao, Rudge, Girgis,
  Kolthammer, Choi, Hong, and Donohoe}}]{xiao07}
\bibinfo{author}{\bibfnamefont{Q.~F.} \bibnamefont{Xiao}},
  \bibinfo{author}{\bibfnamefont{J.}~\bibnamefont{Rudge}},
  \bibinfo{author}{\bibfnamefont{E.}~\bibnamefont{Girgis}},
  \bibinfo{author}{\bibfnamefont{J.}~\bibnamefont{Kolthammer}},
  \bibinfo{author}{\bibfnamefont{B.~C.} \bibnamefont{Choi}},
  \bibinfo{author}{\bibfnamefont{Y.~K.} \bibnamefont{Hong}}, \bibnamefont{and}
  \bibinfo{author}{\bibfnamefont{G.~W.} \bibnamefont{Donohoe}},
  \bibinfo{journal}{J. Appl. Phys.} \textbf{\bibinfo{volume}{102}},
  \bibinfo{pages}{103904} (\bibinfo{year}{2007}).

\bibitem[{\citenamefont{Kravchuk et~al.}(2007)\citenamefont{Kravchuk, Sheka,
  Gaididei, and Mertens}}]{kravchuk07}
\bibinfo{author}{\bibfnamefont{V.~P.} \bibnamefont{Kravchuk}},
  \bibinfo{author}{\bibfnamefont{D.~D.} \bibnamefont{Sheka}},
  \bibinfo{author}{\bibfnamefont{Y.}~\bibnamefont{Gaididei}}, \bibnamefont{and}
  \bibinfo{author}{\bibfnamefont{F.~G.} \bibnamefont{Mertens}},
  \bibinfo{journal}{J. Appl. Phys.} \textbf{\bibinfo{volume}{102}},
  \bibinfo{pages}{043908} (\bibinfo{year}{2007}).

\bibitem[{\citenamefont{Hertel et~al.}(2007)\citenamefont{Hertel, Gliga,
  Fahnle, and Schneider}}]{hertel07}
\bibinfo{author}{\bibfnamefont{R.}~\bibnamefont{Hertel}},
  \bibinfo{author}{\bibfnamefont{S.}~\bibnamefont{Gliga}},
  \bibinfo{author}{\bibfnamefont{M.}~\bibnamefont{Fahnle}}, \bibnamefont{and}
  \bibinfo{author}{\bibfnamefont{C.~M.} \bibnamefont{Schneider}},
  \bibinfo{journal}{Phys. Rev. Lett. (USA)} \textbf{\bibinfo{volume}{98}},
  \bibinfo{pages}{117201/1} (\bibinfo{year}{2007}).

\bibitem[{\citenamefont{Feldtkeller and Thomas}(1965)}]{feldtkeller65}
\bibinfo{author}{\bibfnamefont{E.}~\bibnamefont{Feldtkeller}} \bibnamefont{and}
  \bibinfo{author}{\bibfnamefont{H.}~\bibnamefont{Thomas}},
  \bibinfo{journal}{Phys Kondens Mater} \textbf{\bibinfo{volume}{4}},
  \bibinfo{pages}{8} (\bibinfo{year}{1965}).

\bibitem[{\citenamefont{Raabe et~al.}(2000)\citenamefont{Raabe, Pulwey,
  Sattler, Schweinbock, Zweck, and Weiss}}]{raabe00}
\bibinfo{author}{\bibfnamefont{J.}~\bibnamefont{Raabe}},
  \bibinfo{author}{\bibfnamefont{R.}~\bibnamefont{Pulwey}},
  \bibinfo{author}{\bibfnamefont{R.}~\bibnamefont{Sattler}},
  \bibinfo{author}{\bibfnamefont{T.}~\bibnamefont{Schweinbock}},
  \bibinfo{author}{\bibfnamefont{J.}~\bibnamefont{Zweck}}, \bibnamefont{and}
  \bibinfo{author}{\bibfnamefont{D.}~\bibnamefont{Weiss}}, \bibinfo{journal}{J.
  Appl. Phys.} \textbf{\bibinfo{volume}{88}}, \bibinfo{pages}{4437}
  (\bibinfo{year}{2000}).

\bibitem[{\citenamefont{Wachowiak et~al.}(2002)\citenamefont{Wachowiak, Wiebe,
  Bode, Pietzsch, Morgenstern, and Wiesendanger}}]{wachowiak02}
\bibinfo{author}{\bibfnamefont{A.}~\bibnamefont{Wachowiak}},
  \bibinfo{author}{\bibfnamefont{J.}~\bibnamefont{Wiebe}},
  \bibinfo{author}{\bibfnamefont{M.}~\bibnamefont{Bode}},
  \bibinfo{author}{\bibfnamefont{O.}~\bibnamefont{Pietzsch}},
  \bibinfo{author}{\bibfnamefont{M.}~\bibnamefont{Morgenstern}},
  \bibnamefont{and}
  \bibinfo{author}{\bibfnamefont{R.}~\bibnamefont{Wiesendanger}},
  \bibinfo{journal}{Science} \textbf{\bibinfo{volume}{298}},
  \bibinfo{pages}{577} (\bibinfo{year}{2002}).

\bibitem[{\citenamefont{Chou et~al.}(2007)\citenamefont{Chou, Puzic, Stoll,
  Dolgos, Schutz, Waeyenberge, Vansteenkiste, Tyliszczak, Woltersdorf, and
  Back}}]{chou07}
\bibinfo{author}{\bibfnamefont{K.~W.} \bibnamefont{Chou}},
  \bibinfo{author}{\bibfnamefont{A.}~\bibnamefont{Puzic}},
  \bibinfo{author}{\bibfnamefont{H.}~\bibnamefont{Stoll}},
  \bibinfo{author}{\bibfnamefont{D.}~\bibnamefont{Dolgos}},
  \bibinfo{author}{\bibfnamefont{G.}~\bibnamefont{Schutz}},
  \bibinfo{author}{\bibfnamefont{B.~V.} \bibnamefont{Waeyenberge}},
  \bibinfo{author}{\bibfnamefont{A.}~\bibnamefont{Vansteenkiste}},
  \bibinfo{author}{\bibfnamefont{T.}~\bibnamefont{Tyliszczak}},
  \bibinfo{author}{\bibfnamefont{G.}~\bibnamefont{Woltersdorf}},
  \bibnamefont{and} \bibinfo{author}{\bibfnamefont{C.~H.} \bibnamefont{Back}},
  \bibinfo{journal}{Appl. Phys. Lett.} \textbf{\bibinfo{volume}{90}},
  \bibinfo{pages}{202505} (\bibinfo{year}{2007}).

\bibitem[{\citenamefont{Huber}(1982)}]{huber82}
\bibinfo{author}{\bibfnamefont{D.~L.} \bibnamefont{Huber}},
  \bibinfo{journal}{J. Appl. Phys.} \textbf{\bibinfo{volume}{53}},
  \bibinfo{pages}{1899} (\bibinfo{year}{1982}).

\bibitem[{\citenamefont{Argyle et~al.}(1984)\citenamefont{Argyle, Terrenzio,
  and Slonczewski}}]{argyle84}
\bibinfo{author}{\bibfnamefont{B.~E.} \bibnamefont{Argyle}},
  \bibinfo{author}{\bibfnamefont{E.}~\bibnamefont{Terrenzio}},
  \bibnamefont{and} \bibinfo{author}{\bibfnamefont{J.~C.}
  \bibnamefont{Slonczewski}}, \bibinfo{journal}{1984 Digests Of Intermag '84.
  International Magnetics Conference (cat. No. 84ch1918-2)} pp.
  \bibinfo{pages}{350--350} (\bibinfo{year}{1984}).

\bibitem[{\citenamefont{Guslienko et~al.}(2002)\citenamefont{Guslienko, Ivanov,
  Novosad, Otani, Shima, and Fukamichi}}]{guslienko02}
\bibinfo{author}{\bibfnamefont{K.~Y.} \bibnamefont{Guslienko}},
  \bibinfo{author}{\bibfnamefont{B.~A.} \bibnamefont{Ivanov}},
  \bibinfo{author}{\bibfnamefont{V.}~\bibnamefont{Novosad}},
  \bibinfo{author}{\bibfnamefont{Y.}~\bibnamefont{Otani}},
  \bibinfo{author}{\bibfnamefont{H.}~\bibnamefont{Shima}}, \bibnamefont{and}
  \bibinfo{author}{\bibfnamefont{K.}~\bibnamefont{Fukamichi}},
  \bibinfo{journal}{J. Appl. Phys.} \textbf{\bibinfo{volume}{91}},
  \bibinfo{pages}{8037} (\bibinfo{year}{2002}).

\bibitem[{\citenamefont{Gaididei et~al.}(2008)\citenamefont{Gaididei, Kravchuk,
  Sheka, and Mertens}}]{gaididei08}
\bibinfo{author}{\bibfnamefont{Y.~B.} \bibnamefont{Gaididei}},
  \bibinfo{author}{\bibfnamefont{V.~P.} \bibnamefont{Kravchuk}},
  \bibinfo{author}{\bibfnamefont{D.~D.} \bibnamefont{Sheka}}, \bibnamefont{and}
  \bibinfo{author}{\bibfnamefont{F.~G.} \bibnamefont{Mertens}},
  \bibinfo{journal}{Low Temperature Physics} \textbf{\bibinfo{volume}{34}},
  \bibinfo{pages}{528} (\bibinfo{year}{2008}).

\bibitem[{\citenamefont{Thiaville et~al.}(2003)\citenamefont{Thiaville, Garcia,
  Dittrich, Miltat, and Schrefl}}]{thiaville03}
\bibinfo{author}{\bibfnamefont{A.}~\bibnamefont{Thiaville}},
  \bibinfo{author}{\bibfnamefont{J.~M.} \bibnamefont{Garcia}},
  \bibinfo{author}{\bibfnamefont{R.}~\bibnamefont{Dittrich}},
  \bibinfo{author}{\bibfnamefont{J.}~\bibnamefont{Miltat}}, \bibnamefont{and}
  \bibinfo{author}{\bibfnamefont{T.}~\bibnamefont{Schrefl}},
  \bibinfo{journal}{Phys. Rev., B, Condens, Matter Mater. Phys. (USA)}
  \textbf{\bibinfo{volume}{67}}, \bibinfo{pages}{94410} (\bibinfo{year}{2003}).

\bibitem[{\citenamefont{Novosad et~al.}(2005)\citenamefont{Novosad, Fradin,
  Roy, Buchanan, Guslienko, and Bader}}]{novosad05}
\bibinfo{author}{\bibfnamefont{V.}~\bibnamefont{Novosad}},
  \bibinfo{author}{\bibfnamefont{F.~Y.} \bibnamefont{Fradin}},
  \bibinfo{author}{\bibfnamefont{P.~E.} \bibnamefont{Roy}},
  \bibinfo{author}{\bibfnamefont{K.~S.} \bibnamefont{Buchanan}},
  \bibinfo{author}{\bibfnamefont{K.~Y.} \bibnamefont{Guslienko}},
  \bibnamefont{and} \bibinfo{author}{\bibfnamefont{S.~D.} \bibnamefont{Bader}},
  \bibinfo{journal}{Phys. Rev. B} \textbf{\bibinfo{volume}{72}},
  \bibinfo{pages}{024455} (\bibinfo{year}{2005}).

\bibitem[{\citenamefont{Vansteenkiste et~al.}(2008)\citenamefont{Vansteenkiste,
  Baerdemaeker, Chou, Stoll, Curcic, Tyliszczak, Woltersdorf, Back, Schutz, and
  Waeyenberge}}]{vansteenkiste08}
\bibinfo{author}{\bibfnamefont{A.}~\bibnamefont{Vansteenkiste}},
  \bibinfo{author}{\bibfnamefont{J.~D.} \bibnamefont{Baerdemaeker}},
  \bibinfo{author}{\bibfnamefont{K.~W.} \bibnamefont{Chou}},
  \bibinfo{author}{\bibfnamefont{H.}~\bibnamefont{Stoll}},
  \bibinfo{author}{\bibfnamefont{M.}~\bibnamefont{Curcic}},
  \bibinfo{author}{\bibfnamefont{T.}~\bibnamefont{Tyliszczak}},
  \bibinfo{author}{\bibfnamefont{G.}~\bibnamefont{Woltersdorf}},
  \bibinfo{author}{\bibfnamefont{C.~H.} \bibnamefont{Back}},
  \bibinfo{author}{\bibfnamefont{G.}~\bibnamefont{Schutz}}, \bibnamefont{and}
  \bibinfo{author}{\bibfnamefont{B.~V.} \bibnamefont{Waeyenberge}},
  \bibinfo{journal}{Phys. Rev. B} \textbf{\bibinfo{volume}{77}},
  \bibinfo{pages}{144420} (\bibinfo{year}{2008}).

\bibitem[{\citenamefont{Kilcoyne et~al.}(2003)\citenamefont{Kilcoyne,
  Tyliszczak, Steele, Fakra, Hitchcock, Franck, Anderson, Harteneck, Rightor,
  Mitchell et~al.}}]{kilcoyne03}
\bibinfo{author}{\bibfnamefont{A.}~\bibnamefont{Kilcoyne}},
  \bibinfo{author}{\bibfnamefont{T.}~\bibnamefont{Tyliszczak}},
  \bibinfo{author}{\bibfnamefont{W.}~\bibnamefont{Steele}},
  \bibinfo{author}{\bibfnamefont{S.}~\bibnamefont{Fakra}},
  \bibinfo{author}{\bibfnamefont{P.}~\bibnamefont{Hitchcock}},
  \bibinfo{author}{\bibfnamefont{K.}~\bibnamefont{Franck}},
  \bibinfo{author}{\bibfnamefont{E.}~\bibnamefont{Anderson}},
  \bibinfo{author}{\bibfnamefont{B.}~\bibnamefont{Harteneck}},
  \bibinfo{author}{\bibfnamefont{E.}~\bibnamefont{Rightor}},
  \bibinfo{author}{\bibfnamefont{G.}~\bibnamefont{Mitchell}},
  \bibnamefont{et~al.}, \bibinfo{journal}{Journal of Synchrotron Radiation}
  \textbf{\bibinfo{volume}{10}}, \bibinfo{pages}{125} (\bibinfo{year}{2003}).

\bibitem[{\citenamefont{Sch{\"u}tz et~al.}(1987)\citenamefont{Sch{\"u}tz,
  Wagner, Wilhelm, Kienle, Zeller, Frahm, and Materlik}}]{schutz87}
\bibinfo{author}{\bibfnamefont{G.}~\bibnamefont{Sch{\"u}tz}},
  \bibinfo{author}{\bibfnamefont{W.}~\bibnamefont{Wagner}},
  \bibinfo{author}{\bibfnamefont{W.}~\bibnamefont{Wilhelm}},
  \bibinfo{author}{\bibfnamefont{P.}~\bibnamefont{Kienle}},
  \bibinfo{author}{\bibfnamefont{R.}~\bibnamefont{Zeller}},
  \bibinfo{author}{\bibfnamefont{R.}~\bibnamefont{Frahm}}, \bibnamefont{and}
  \bibinfo{author}{\bibfnamefont{G.}~\bibnamefont{Materlik}},
  \bibinfo{journal}{Phys. Rev. Lett.} \textbf{\bibinfo{volume}{58}},
  \bibinfo{pages}{737} (\bibinfo{year}{1987}).

\bibitem[{\citenamefont{Landau and Lifshitz}(1935)}]{landau35}
\bibinfo{author}{\bibfnamefont{L.}~\bibnamefont{Landau}} \bibnamefont{and}
  \bibinfo{author}{\bibfnamefont{E.}~\bibnamefont{Lifshitz}},
  \bibinfo{journal}{Phys. Z. Sowietunion} \textbf{\bibinfo{volume}{8}}
  (\bibinfo{year}{1935}).

\bibitem[{\citenamefont{Visscher and Apalkov}(2004)}]{visscher04}
\bibinfo{author}{\bibfnamefont{P.~B.} \bibnamefont{Visscher}} \bibnamefont{and}
  \bibinfo{author}{\bibfnamefont{D.~M.} \bibnamefont{Apalkov}},
  \bibinfo{journal}{Physica B} \textbf{\bibinfo{volume}{343}},
  \bibinfo{pages}{184} (\bibinfo{year}{2004}).

\end{thebibliography}

\section*{Acknowledgements}
Financial support by The Institute for the promotion of Innovation by Science and
Technology in Flanders (IWT-Flanders) and by the Research Foundation Flanders (FWO-Flanders)
through the research grant 60170.06 are gratefully acknowledged. The Advanced Light Source
is supported by the Director, Office of Science, Office of Basic Energy Sciences, of the
U.S. Department of Energy.

\section*{Author contributions}

Analyses and micromagnetic simulations: A.V.; Experiments: A.V., K.W.C., M.W., M.C., V.S., H.S., T.T.;  Writing the paper: A.V., K.W.C., G.W., C.H.B., H.S., B.V.; Sample preparation: G.W., C.H.B.; Project planning: H.S., B.V., G.S.

\end{document}